\documentclass[twocolumn]{aastex}
\usepackage{amsmath}
\usepackage{amsbsy}

\usepackage{booktabs}

\shorttitle{Active Region Irradiance During Quiescent Periods}

\shortauthors{Kazachenko et al.}

\begin{document}
\title{Active Region Irradiance During Quiescent Periods: New Insights from Sun-as-a-star spectra}
\author{Maria~D.~Kazachenko\altaffilmark{1,2}, Hugh Hudson\altaffilmark{3,4}}
\altaffiltext{1}{Department of Astrophysical and Planetary Sciences, University of Colorado Boulder, 2000 Colorado Ave, Boulder, CO 80305, USA}
\altaffiltext{2}{National Solar Observatory, University of Colorado Boulder, 3665 Discovery Drive, Boulder, CO 80303, USA}
\altaffiltext{3}{Space Sciences Laboratory, University of California--Berkeley, Berkeley, CA 94720, USA}
\altaffiltext{4}{SUPA School of Physics \& Astronomy, University of Glasgow, UK}
\email{maria.kazachenko@colorado.edu}
\begin{abstract}
How much energy do solar active regions (ARs) typically radiate during quiescent periods?  This is a fundamental question for storage and release models of flares and ARs, yet it is presently poorly answered by observations.  Here we use the ``Sun-as-a-point-sourceÕÕ spectra from the EUV Variability Experiment (EVE) on the Solar Dynamics Observatory (SDO) to provide a novel estimate of radiative energy losses of an evolving active region.  Although EVE provides excellent spectral ($5-105$ nm) and temperature ($2-25$  MK) coverage for AR analysis, to our knowledge, these data have not been used for this purpose due to the lack of spatial resolution and the likelihood of source confusion. Here we present a way around this problem. We analyze EVE data time series, when only one large AR 11520 was present on the disk.  By subtracting the quiet Sun background, we estimate the radiative contribution in EUV from the AR alone.  We estimate the mean AR irradiance and cumulative AR radiative energy losses in the $1-300$~\AA\  and astronomical standard ROSAT-PSPC, $3-124$~\AA, passbands and compare these to the magnetic energy injection rate through the photosphere, and to variations of the solar cycle luminosity. We find that while AR radiative energy losses are ~100 times smaller than typical magnetic energy injection rates at the photosphere, they are an order of magnitude larger or similar to the bolometric radiated energies associated with large flares. This study is the first detailed analysis of AR thermal properties using EVE Sun-as-a-star observations, opening doors to AR studies on other stars. 
\end{abstract}


%
%
\section{Introduction}

The evolution of an active region (AR) in the solar corona reflects the evolution of the photospheric magnetic fields connected to it (see the review by \citealt{Driel2015}).  
Many multi-wavelength analyses have studied the long-term evolution of ARs, including the behavior of the magnetic field and its non-potentiality (e.g. \citealt{Demoulin2002}), flare and CME activity (e.g. \citealt{Iglesias2019}), helicity budget  (e.g. \citealt{Liu2012}), coronal heating  (e.g. \citealt{Warren2012,Reva2018}), coronal intensity, temperature and emission measure (e.g. \citealt{Fisher1998,Tripathi2011,Ugarte2017}), total irradiance (e.g. \citealt{Ortiz2004,Zahid2004}). 

While many aspects of long-term active region evolution have been well studied \citep{Driel2015}, we are still missing  quantitative observational estimates of how much energy ARs lose to radiation as they evolve. 
Specifically, how much of the magnetic energy injected as Poynting flux through the photosphere is lost due to radiation?  
This estimate is essential to constrain the total energy budget of the AR. 
In principle the Poynting flux should match the budget of magnetic energy stored in the AR and its radiative (and other) energy losses. 
In practice, however, most of the current approaches estimate AR total magnetic energy as a cumulative sum of the photospheric Poynting fluxes \citep{Kazachenko2015,Liu2012,Lumme2019,Tziotziou2013ApJ,Vemareddy2015}, neglecting radiative energy losses.  

Deriving radiative energy losses from the observations requires broad spectral measurements covering a range of characteristic temperatures. Here the Extreme Ultraviolet Variability (EUV) Experiment (EVE; \citealt{Woods2012}) on-board Solar Dynamics Observatory (SDO; \citealt{Pesnell2012}) has ideal properties spectroscopically.
It acquires continuous full-disk EUV spectra of Sun-as-a-star in the range of $5$ to $105$~nm every $10$ seconds with spectral resolution of $0.1$~nm and an excellent signal-to-noise ratio.
The MEGS-A component of EVE in particular contains a wide variety of iron emission lines formed around $10^{5.7}-10^{6.8}$~K  ($0.5-6.3$ MK), 
making it an excellent basis for studying solar temperature profiles during both active and quiescent periods.  

Although EVE provides full spectral coverage for AR analysis, to our knowledge, no EVE data have been used explicitly for AR analysis. 
The main reason for this is lack of spatial resolution in EVE observations -- since EVE observes the Sun-as-a-star spectrum, it cannot simply distinguish between AR and quiet-Sun contributions. However, when only one AR is present on the disk we can subtract the Sun-as-a-star spectrum when no ARs are present (the quiet-Sun spectrum), to derive the spectrum of AR alone. 
In this paper, we use such EVE difference spectra to first find the differential emission measure (DEM) of the AR and then use it to quantify AR radiative energy losses as a function of time in the $1-300$~\AA{} spectral range.

The Sun-as-a-star approach here explicitly relates to the solar irradiance at one AU, equivalent to an astronomical ``spectral energy distribution'' or SED. The standard solar reference work for active-region luminosity remains the classical review of \cite{Withbroe1977}, which focused on assessing the luminosity as a surface brightness, as related to classical 1D semi-empirical model atmospheres. 
Our approach here results in the estimates of the luminosity, $L_\mathrm{AR}$, of an active region during its disk passage, which we compare with the total unsigned magnetic flux of the region \citep[cf.][]{Fisher1998} and the magnetic energy flux through the photosphere, the Poynting flux. 

The paper is organized as follows. In Section~\ref{data}, we describe the EVE data during a passage of AR 11520 and the analysis we used to find fluxes in lines of interest. 
In Section~\ref{methods}, we describe the methods we use to find AR Differential Emission Measures (DEMs) and AR radiative losses. 
In Section~\ref{results} we present DEMs from EVE and compare them with DEMs derived from the averaged intensities from the Atmospheric Imaging Assembly  \citep[AIA;][]{Lemen2012} data; we show how the region's DEM evolves as a function of time, derive its radiative losses and total radiated energy and compare our results with  previous scaling relationship between AR X-ray emission and magnetic flux \citep{Fisher1998}. 
Finally, in Section~\ref{discussion}, we summarize our results and draw conclusions, including comparisons with the classical reference of \cite{Withbroe1977}.

\begin{figure*}[htb!]
  \centering  


\resizebox{0.32\hsize}{!}{\includegraphics[angle=0,trim=0.35cm 1.8cm 1.6cm 2.4cm,clip,width=\textwidth]{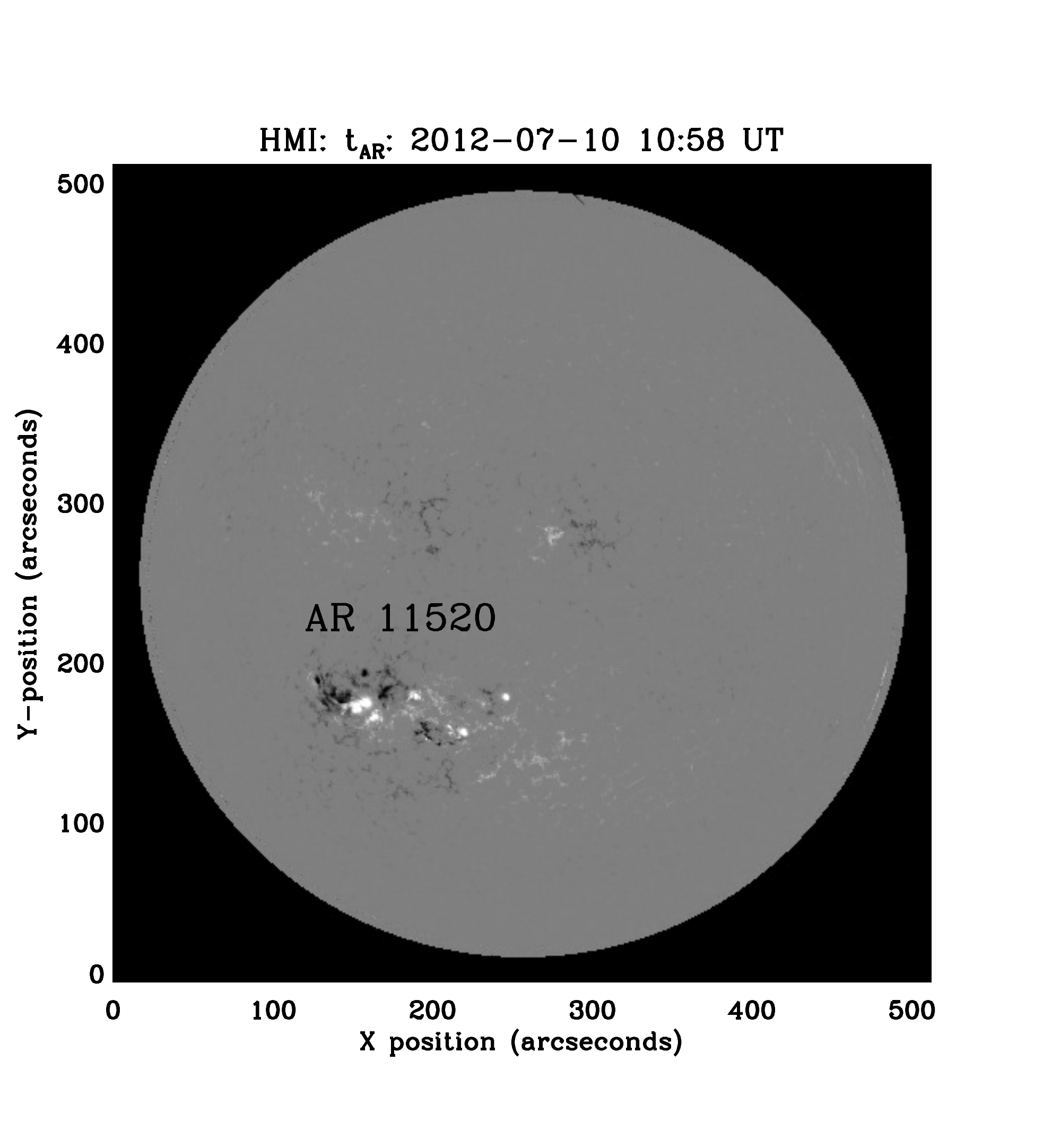}} 
\resizebox{0.32\hsize}{!}{\includegraphics[angle=0,trim=0.35cm 1.8cm 1.6cm 2.4cm,clip,width=\textwidth]{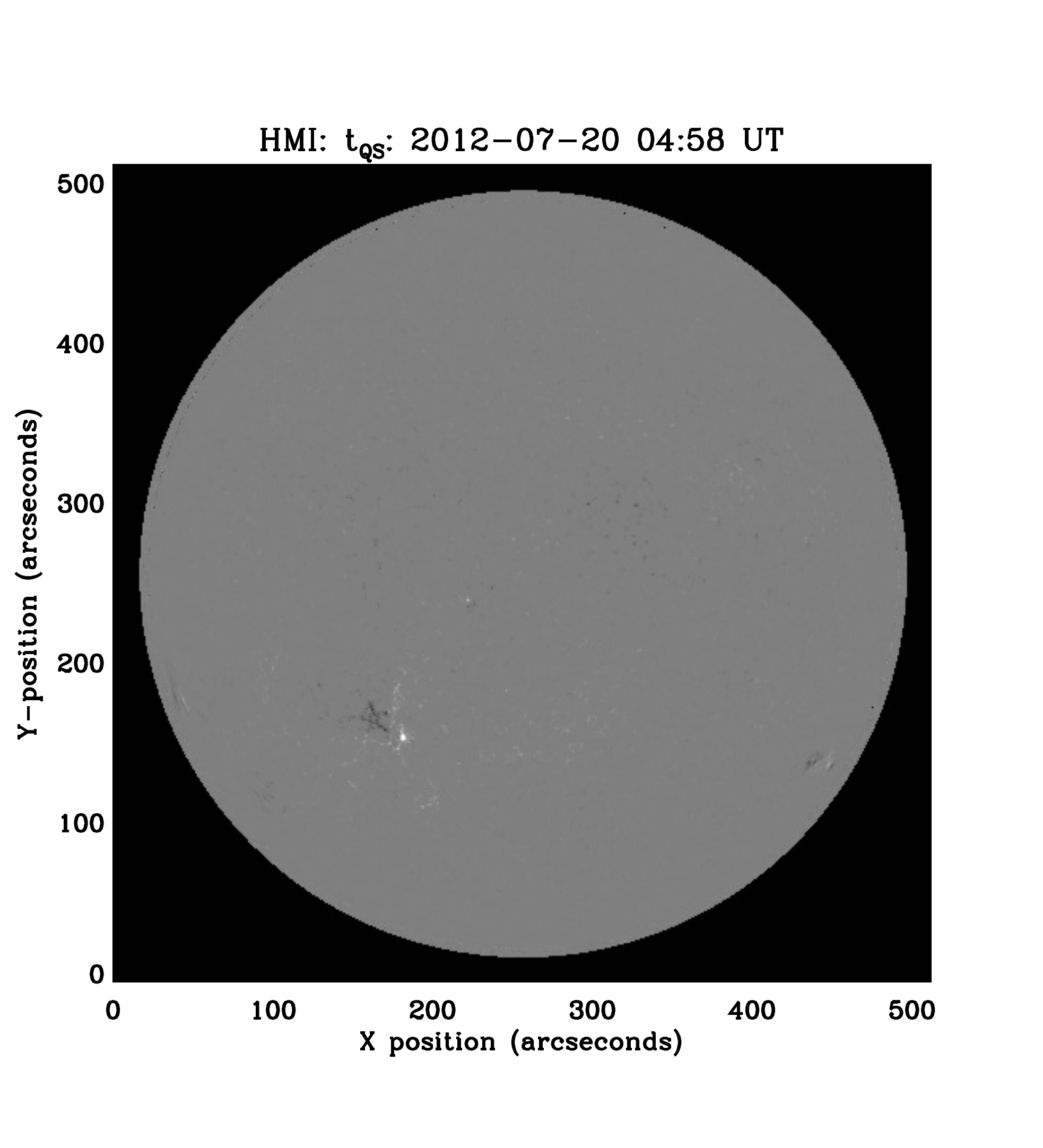}}  
\resizebox{0.3\hsize}{!}{\includegraphics[angle=0]{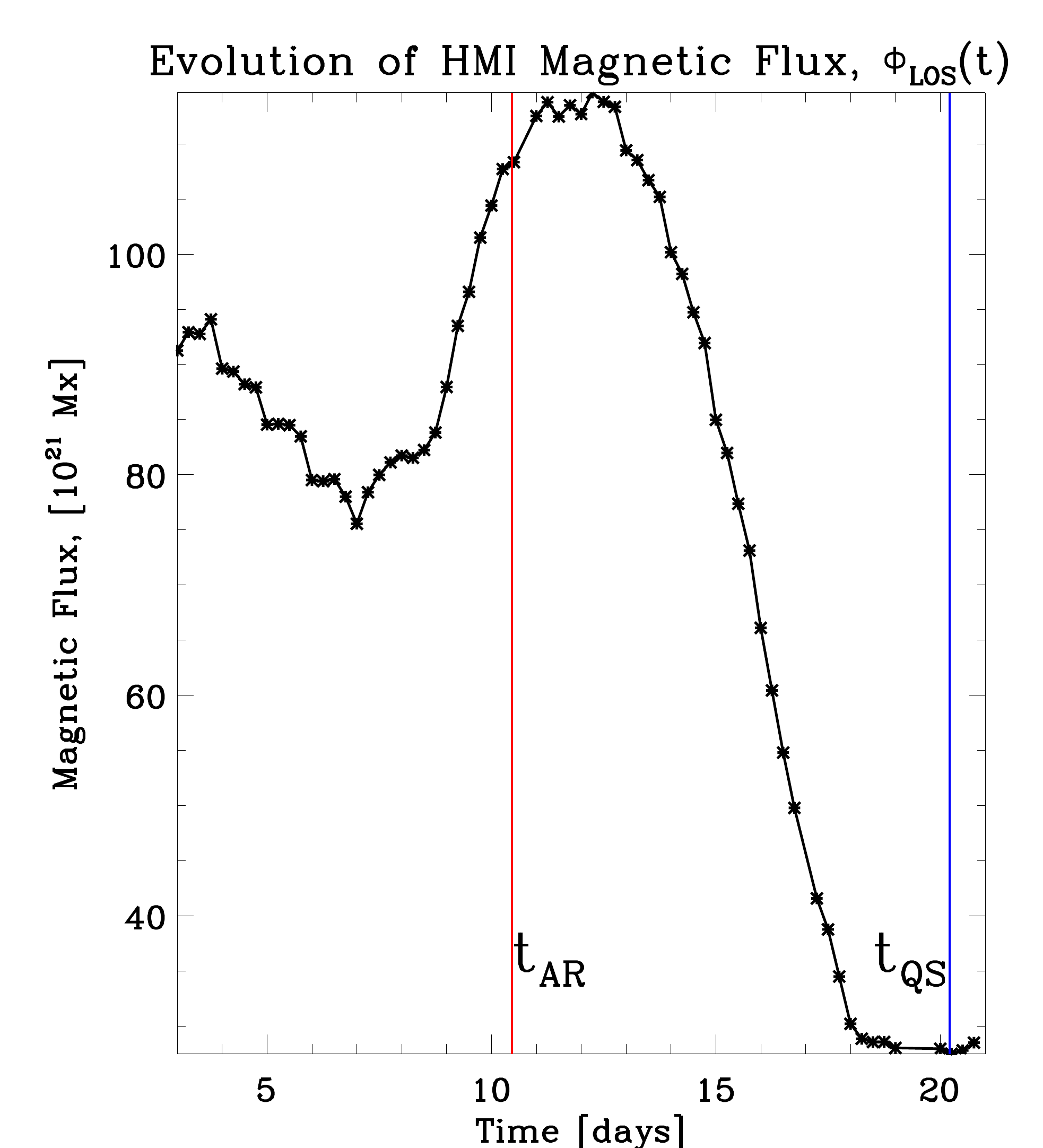}}
\resizebox{0.32\hsize}{!}{\includegraphics[angle=0,trim=0.05cm 1.55cm 1.3cm 2.0cm,clip,width=\textwidth]{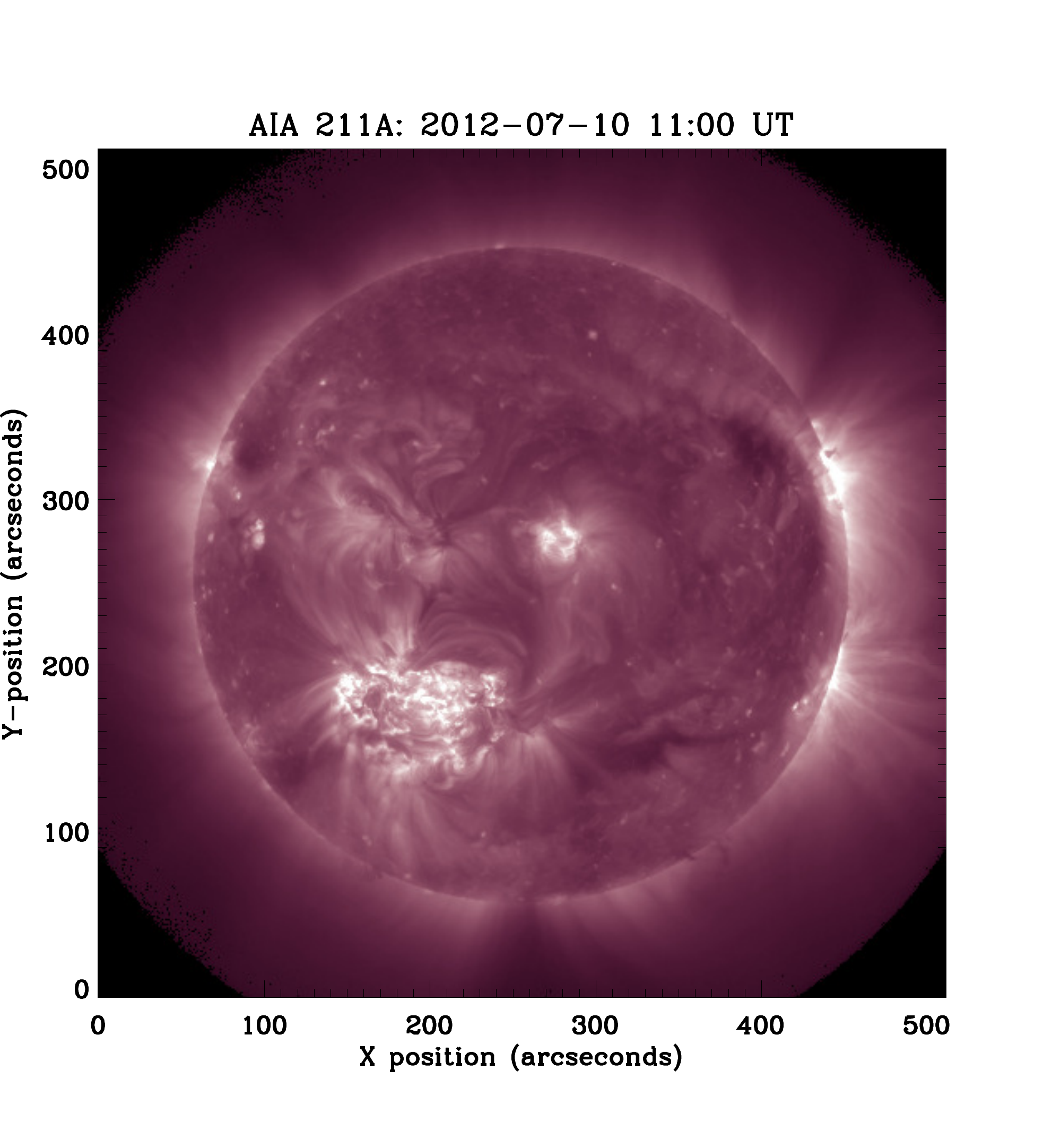}} 
\resizebox{0.32\hsize}{!}{\includegraphics[angle=0,trim=0.05cm 1.55cm 1.3cm 2.0cm,clip,width=\textwidth]{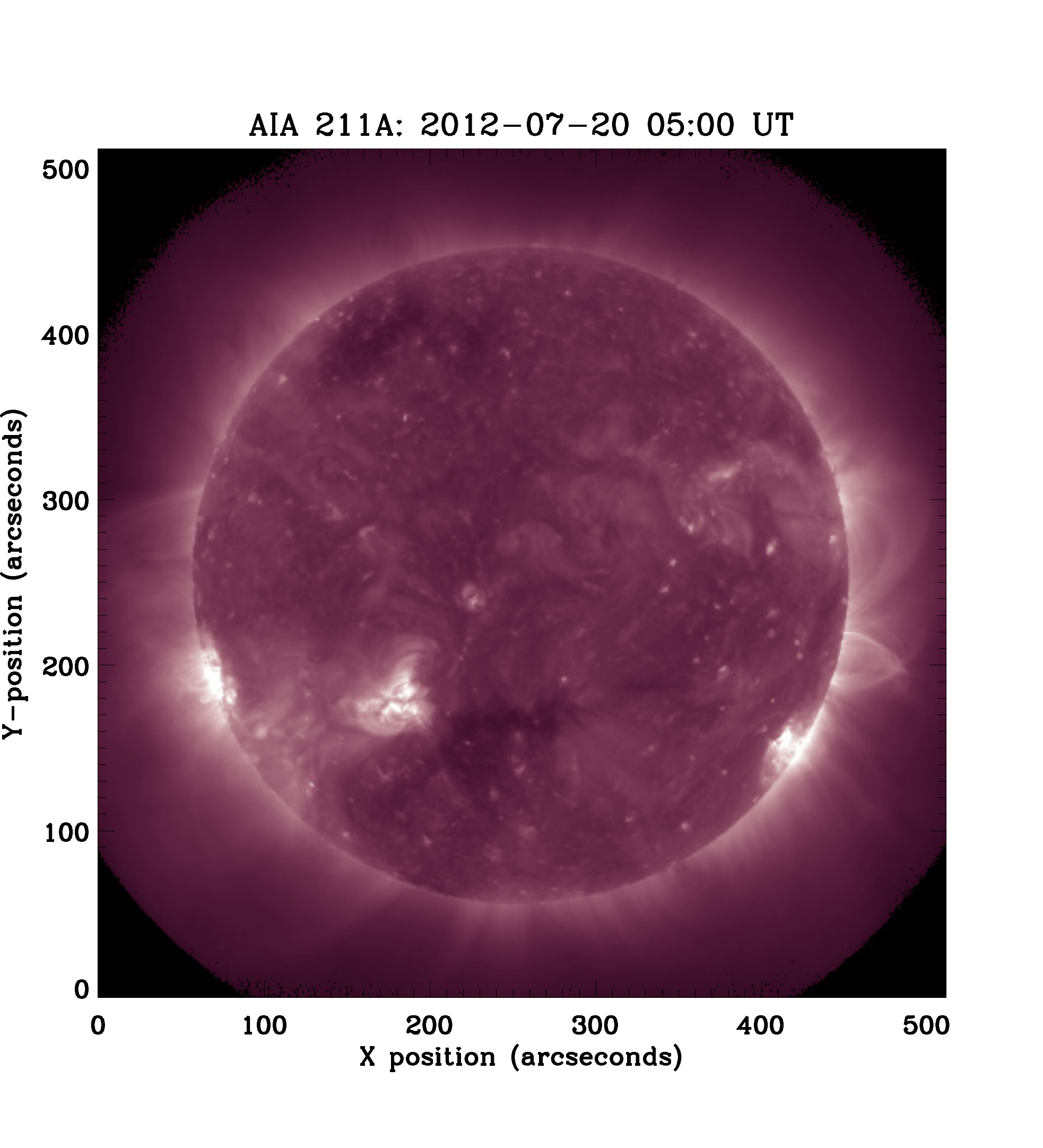}} 
\resizebox{0.3\hsize}{!}{\includegraphics[angle=0]{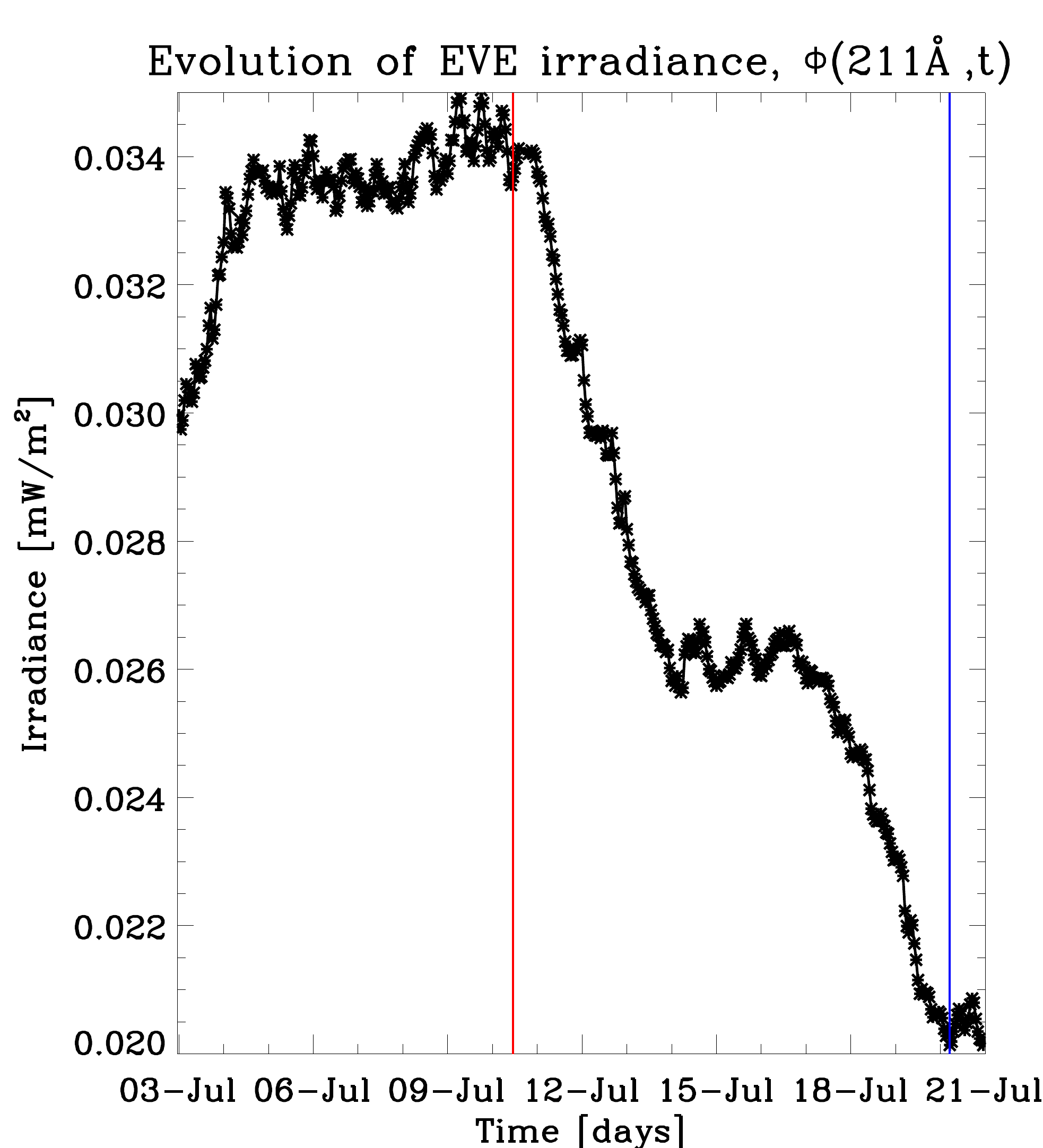}}
  \caption{AR 11520:  HMI LOS magnetic field ({\it Top}) and AIA $211 $~\AA{}  maps ({\it Bottom}) around AR and quiet Sun times, $t_\mathrm{AR}$=10-Jul 10:59 UT ({\it Left}) and $t_\mathrm{QS}$=20-Jul 04:59 UT ({\it Middle}), respectively. 
  The {\it Right} column shows the evolution of HMI disk-integrated unsigned LOS magnetic flux, $\Phi_\mathrm{LOS}(t)$, and Fe{\sc xiv} 211~\AA{} EVE irradiance (line integrated spectrum), $\Phi(211$\AA{}$,t)$, as AR 11520 crossed the disk. 
  The vertical red and blue dotted lines show $t_\mathrm{AR}$ and $t_\mathrm{QS}$, respectively. See \S\ref{data}.}
  \label{disk}
\end{figure*}

 \begin{figure*}[tb!]
  \centering 
\resizebox{1.05\hsize}{!}{\includegraphics[angle=0,trim=0.0cm 0cm 0cm 0cm,clip,width=\textwidth]{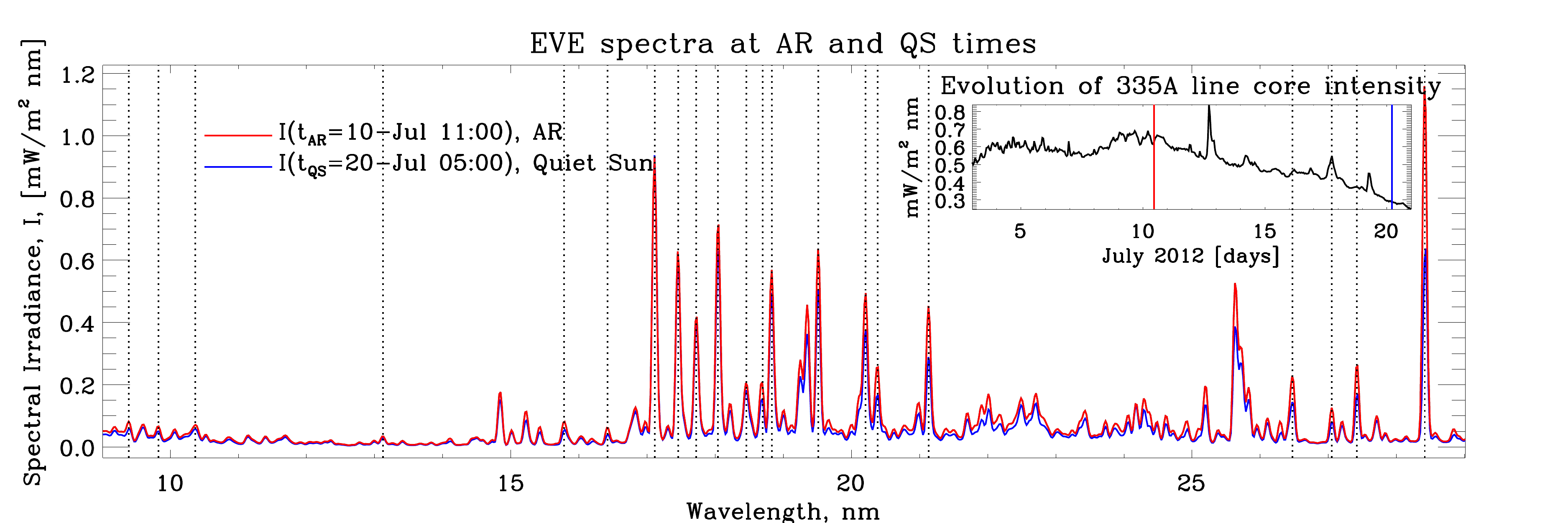}}
\resizebox{1.05\hsize}{!}{\includegraphics[angle=0]{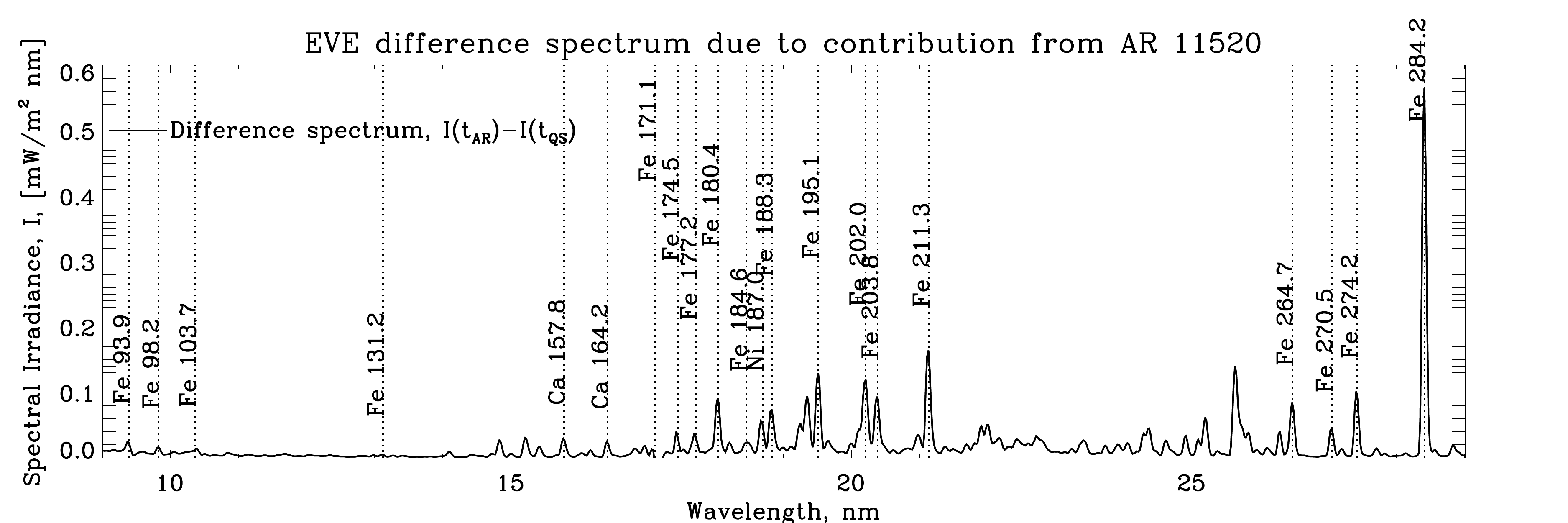}}
  \caption{{\it Top:} EVE Sun-as-a-star spectra (spectral irradiances) at AR and QS times, $I(\lambda,t_\mathrm{AR})$ and $I(\lambda,t_\mathrm{QS})$, repectively (red and blue colors); {\it Top right} panel shows evolution of the brightest FeXVI~$335$~\AA{} line core intensity, $I(\lambda=335.45$~\AA{}$,t)$, from July 3 to July 21; {\it Bottom:} EVE difference spectrum due to contribution from the active region alone at $t_\mathrm{AR}$, $I(t_\mathrm{AR})-I(t_\mathrm{QS})$. 
  Dotted lines show spectral lines selected for the DEM analysis. See the line list in Table~\ref{tab:line_power} and more details in \S\ref{data}.} 
  \label{spec}
\end{figure*}

\begin{table} \small \caption{The $24$ EVE lines and a subset of $6$ AIA target lines (marked with *) initially selected for the AR analysis. Due to negative irradiance during AR evolution we further exclude $171.11$ \AA{} line from our analysis, resulting in a final list of $23$ EVE and $5$ AIA lines. Columns correspond to each line wavelength, ion, temperature of the peak of the contribution function ($T_\mathrm{max}$) and the maximum value of the contribution function ($G_\mathrm{max}$). See \S\ref{data}.} \small \begin{center}
\begin{tabular}{cllcc}
i & Line & Ion & $Log_\mathrm{10}(T_\mathrm{max})$ & $G_\mathrm{max}$, $10^{-24}$ \\
 & \AA{}  &     & K           & ergs cm$^{-3}$s$^{-1}$ \\
\hline
       0 & 131.24* & FeVIII & 5.70* &  2.00 \\
       1 & 171.11* & FeIX & 5.80* & 49.47 \\
       2 & 174.54 & FeX & 6.00 & 18.33 \\
       3 & 177.23 & FeX & 6.00 & 10.28 \\
       4 & 184.57 & FeX & 6.00 &  4.72 \\
       5 & 186.98 & NiXI & 6.10 &  0.10 \\
       6 & 180.43 & FeXI & 6.10 & 11.98 \\
       7 & 188.30 & FeXI & 6.10 &  9.04 \\
       8 & 195.13* & FeXII & 6.10* & 12.40 \\
       9 & 202.04 & FeXIII & 6.20 &  3.79 \\
      10 & 203.83 & FeXIII & 6.20 & 12.99 \\
      11 & 211.33* & FeXIV & 6.30* &  6.03 \\
      12 & 264.72 & FeXIV & 6.30 &  5.76 \\
      13 & 270.52 & FeXIV & 6.30 &  2.28 \\
      14 & 274.20 & FeXIV & 6.30 &  3.14 \\
      15 & 284.16 & FeXV & 6.30 & 25.76 \\
      16 & 335.45* & FeXVI & 6.40* & 12.03 \\
      17 & 360.81 & FeXVI & 6.40 &  5.78 \\
      18 &  69.68 & FeXV & 6.40 &  5.78 \\
      19 & 157.82 & CaXVI & 6.70 &  0.02 \\
      20 & 164.17 & CaXVI & 6.70 &  0.06 \\
      21 &  93.93* & FeXVIII & 6.80* &  1.27 \\
      22 & 103.70 & FeXVIII & 6.80 &  0.42 \\
      23 &  98.25 & FeXVII & 6.80 &  0.42 \\
\end{tabular} \normalsize  \normalsize \label{tab:line_power} \end{center} \end{table}

\section{Data: EVE Observations of AR 11520}\label{data}

We selected AR 11520 for this study, as shown in ~Figure~\ref{disk}.  
This region was the only AR present on the solar disk from its central meridian crossing ($\sim$11~July~2012) to its disappearance over the East limb, allowing us to attribute the disk-integrated properties to this region over this interval.
Specifically, AR 11520 appeared on the East limb on $t_\mathrm{East}$~=~05-Jul-2012~(S15E70), crossed the central meridian on $t_\mathrm{CM}$~=~12-Jul-2012~(S16W08) and disappeared behind the West limb on $t_\mathrm{West}$~=~18-Jul-2012~(S17W89).  
The Sun did not have any large ARs on July 21 2012, hence we chose $t_\mathrm{QS}$~=~20-Jul-2012~05:00 UT as the quiet-Sun reference time. 
We chose $t_\mathrm{AR}$~=~10-Jul-2012~11:00 UT as the primary sample time, when the AR was $15^\circ$ East from the central meridian; $t_\mathrm{AR}$ corresponded to a quiescent period of AR evolution one hour after a C2.0-class flare.  As AR 11520 crossed the disk, it hosted one X1.4 flare on 12 July 2012, four M-class flares (M1.1, M2.0, M1.7, M1.0) and 37 C-class flares. 

Figure~\ref{disk} illustrates the temporal evolution of image-resolved magnetic field and EUV 211~\AA{} irradiance: the left and middle columns show line-of-sight (LOS) magnetic field maps from Helioseismic and Magnetic Imager (HMI, \citealt{Schou2012}) and 211~\AA{} images from AIA at $t_\mathrm{AR}$ and $t_\mathrm{QS}$, respectively. 
The right column shows the evolution of disk-integrated LOS magnetic flux, $\Phi_\mathrm{LOS}(t)$, and the 211~\AA{} irradiance, $\Phi(211$\AA{}$,t)$, over the $17$-day interval, with red and blue vertical lines marking $t_\mathrm{AR}$ and $t_\mathrm{QS}$, respectively. For consistent comparison with irradiances, we did not account for magnetic field inclination nor foreshortening in $\Phi_\mathrm{LOS}(t)$ calculation. We also ignored small contributions into  $\Phi_\mathrm{LOS}(t)$ and EVE irradiances from minor regions after $t_\mathrm{AR}$, assuming that these originate from AR 11520. At the beginning of AR disk passage (before $t_\mathrm{AR}$), other preceding regions dominate.
The smooth variation of the LOS magnetic flux around the central meridian passage shows the dominance of this region and is consistent with the expected 
nearly-vertical field orientation.
Taking the difference of magnetic fluxes at the AR and QS times, we evaluate the target region's LOS unsigned magnetic flux at $t_\mathrm{AR}$, $\Phi_\mathrm{LOS,AR}(t_\mathrm{AR})=\Phi_\mathrm{LOS}(t_\mathrm{AR})-\Phi_\mathrm{LOS}(t_\mathrm{QS})=8\times10^{22}$~Mx. 
Alternatively, using a Mercator-deprojected HMI vector magnetic field map that includes the AR alone, we find a more accurate vertical (radial) unsigned magnetic flux $\Phi_\mathrm{r,AR}=11.8\times10^{22}$~Mx.
Comparing this estimate with magnetic flux distributions from other known regions (see Figure 8, \citealt{Kazachenko2017}), we characterize AR 11520 as a large region, with $\Phi_\mathrm{AR}$ exceeding that of $80\%$ of all solar active regions (for comparison the well-studied region AR 11158 had about 1/3 as much unsigned magnetic flux, $4\times10^{22}$~Mx).


To track the region's evolution across the EUV spectrum, we used EVE spectral irradiances, $I(\mathrm{\lambda},t)$,  from $t_\mathrm{start}$~=~03-Jul-12~09:00~UT  to $t_\mathrm{end}$~=~20-Jul-12~00:00~UT. 
We chose an hourly cadence and derived $429$ spectra ranging from $6.5$ to $37$ nm.  
When an active region passes across the solar disk, an enhancement is observed across the EUV spectrum. 
We used this enhancement to derive the {\it difference spectrum}, comparing $t_\mathrm{AR}$ to $t_\mathrm{QS}$  and thus isolating the target region alone; we can thus study its properties independent of the quiet-Sun background. 
This approach is not perfect because the presence of the region may itself alter the background spectrum; for this reason we have restricted our analysis to optically thin lines where this problem may not be significant.

Figure~\ref{spec} shows the EVE spectrum (spectral irradiance) at $t_\mathrm{AR}$ and  $t_\mathrm{QS}$,  namely $I(\mathrm{\lambda},t_\mathrm{AR})$ and $I(\mathrm{\lambda},t_\mathrm{QS})$ (top), and the difference spectrum due in principle to AR 11520 contribution alone at $t_\mathrm{AR}$, $I(\mathrm{\lambda},t_\mathrm{AR})-I(\mathrm{\lambda},t_\mathrm{QS})$ (bottom). 
The dotted lines correspond to $24$ selected spectral lines, $\{\lambda_i\}$, that we chose to describe the EUV spectrum (see Table~\ref{tab:line_power} for a line list). 

\begin{figure*}[htb!]
  \centering 
\resizebox{1.0\hsize}{!}{\includegraphics[angle=0]{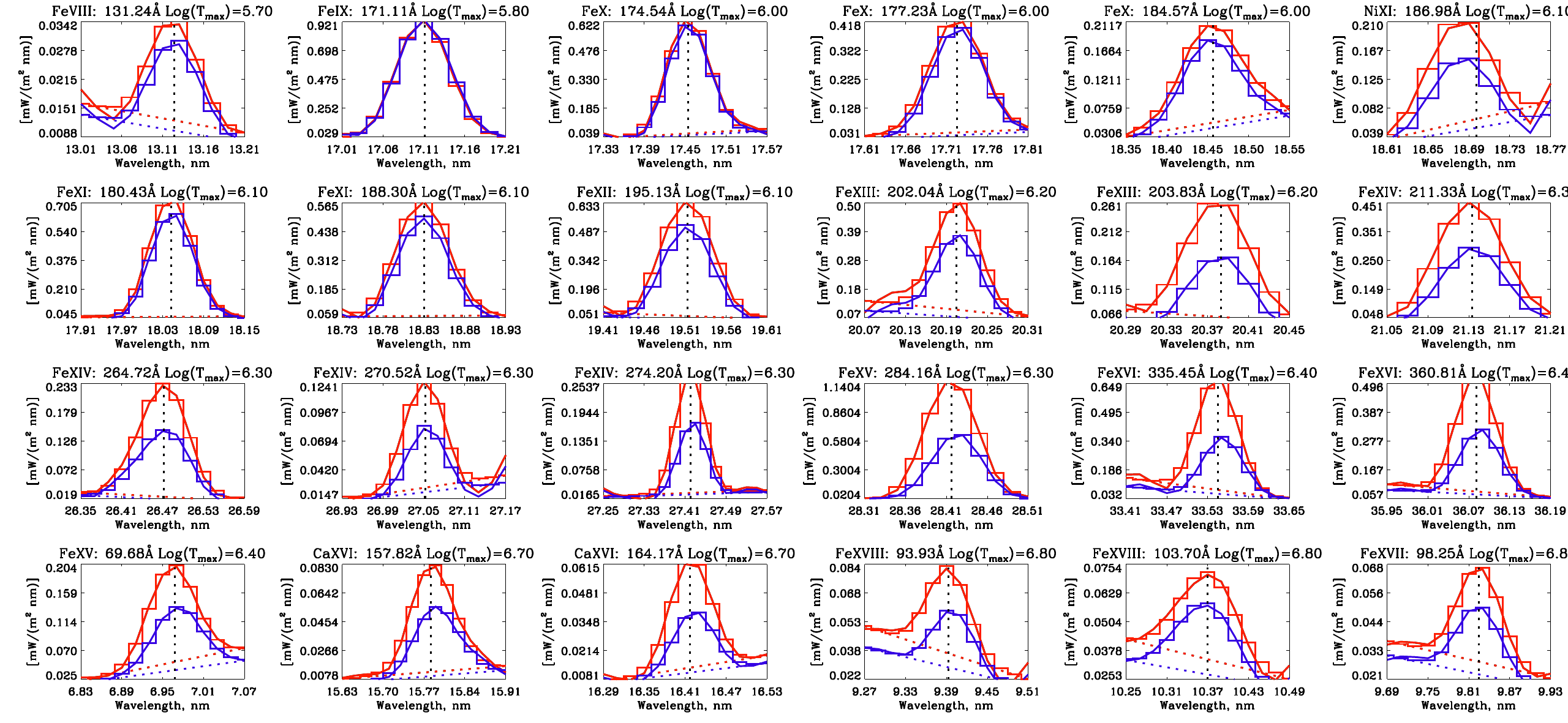}}
 \caption{EVE line profiles (solid curves) with corresponding Gaussian fits at AR ($t_\mathrm{AR}$=10-Jul 10:59 UT, blue) and quiet-Sun ($t_\mathrm{QS}$=20-Jul 04:59 UT, blue) times. Vertical and inclined dotted lines show calculated line center location and the background value, respectively. See \S\ref{data}.}
  \label{linefits}
\end{figure*}

To quantify the total flux in each line, taking into account possible line shifts and background offsets, we fitted the curve of each line with a Gaussian function. Figure~\ref{linefits} shows an example of Gaussian fits for $24$ EVE spectral lines at $t_\mathrm{AR}$ and  $t_\mathrm{QS}$, respectively. We then integrated the spectral irradiance below each Gaussian around the line center to estimate the net irradiance corresponding to each spectral line at time $t$: 
\begin{equation}
\begin{aligned}
   \Delta \Phi(\mathrm{\lambda_i,t}) &=\int_{\lambda_i} I(\mathrm{\lambda,t}) d\mathrm{\lambda}-\int_{\lambda_i} I(\mathrm{\lambda,t_{QS}})\\
   &=\Phi(\lambda_i,t)-\Phi(\lambda_i,t_\mathrm{QS}).
\end{aligned}\label{eq_flux}
\end{equation}

\begin{figure*}[htb!]
  \centering 
\resizebox{1.0\hsize}{!}{\includegraphics[angle=0]{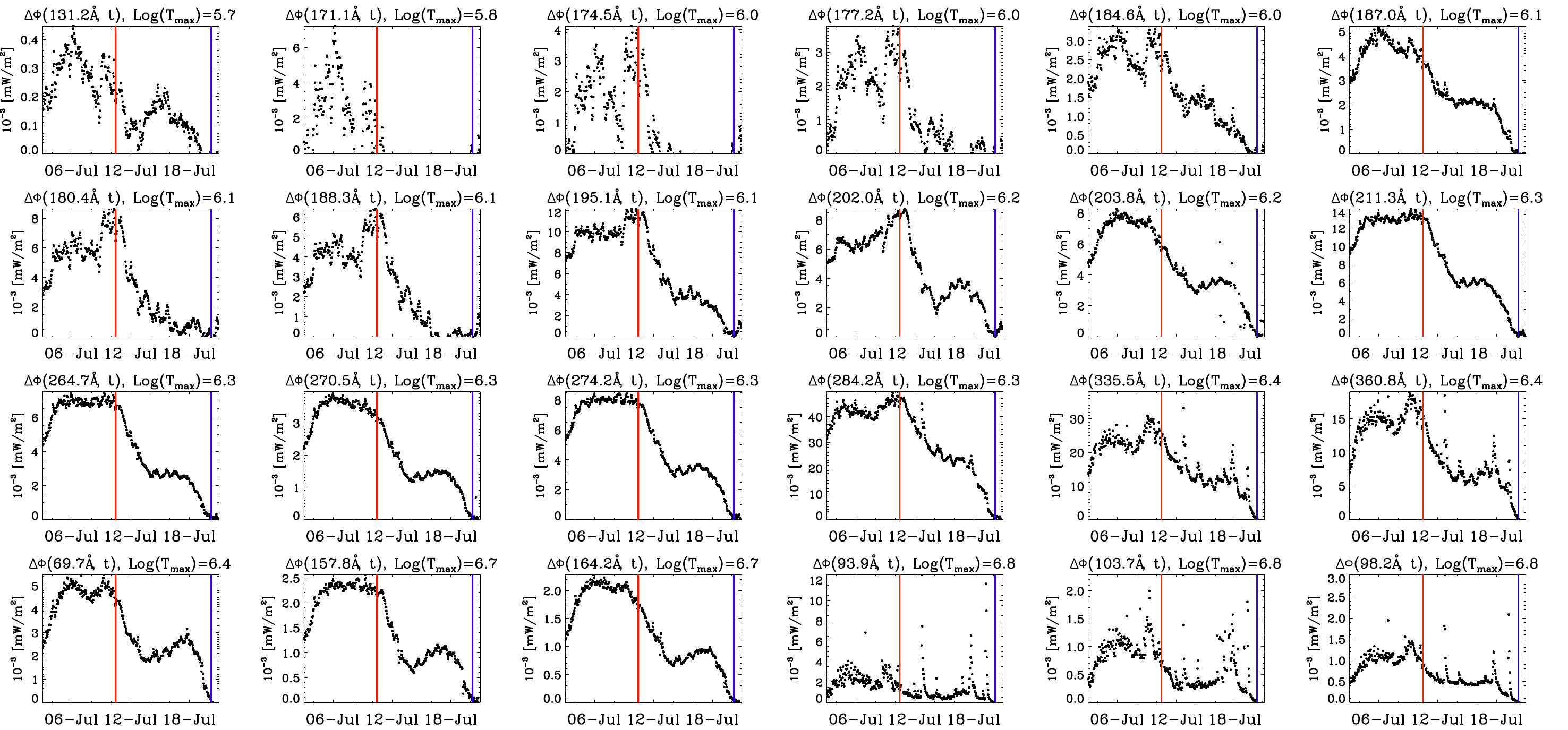}}
 \caption{Evolution of AR 11520 irradiances (line-integrated EVE spectra), $\Delta \Phi(\mathrm{\lambda_i,t})$, from July 3 to July 21 for $24$ EVE lines derived from difference spectrum with respect to the quiet-Sun (see Eq.~\ref{eq_flux}). Vertical red and blue lines show $t_\mathrm{AR}$ and $t_\mathrm{QS}$, respectively. See \S\ref{data}.}
  \label{linefits_evol}
\end{figure*}

Figure~\ref{linefits_evol} shows the result of this fitting: evolution of irradiances, $\Delta \Phi(\mathrm{\lambda_i,t})$, in $24$ EVE lines corresponding to AR 11520 as it crossed the disk. Note, how $171.1$~\AA{} line irradiance dropped below zero as AR 11520 crossed the disk (Panel 2, following the red line). 
These negative values are caused by ``dark canopies'' or ``circumfacular regions'' around the AR, which are darker that the quiet Sun in $171.1$~\AA{} line.
Originally discovered by \citet{1903PYerO...3....1H} as a chromospheric effect, they also appear clearly in the nominally coronal 171~\AA~band \citep{Wang2011}, but without a detailed physical explanation yet.
Since negative irradiances complicate the analysis, we exclude the $171.1$~\AA{} line from our further EVE data analysis, resulting in a final list of 23 EVE and 5 AIA lines \citep{1903PYerO...3....1H}.

\section{Methods: finding the DEM and radiated energy using EVE data}\label{methods}

The flux differences we derive from EVE for each line, $\Delta \Phi(\mathrm{\lambda_i,t})$, are related to emissions as expressed in terms of the contribution function $G(\lambda_i,T_e,N_e)$,  and the differential emission measure, $DEM (T_e,t)$, by 
\begin{equation}
\Delta \Phi(\mathrm{\lambda_i,t})=\frac{A_\mathrm{AR}}{4\pi R^2}  \int_{T_e} G(\lambda_i,T_e, N_e) DEM (T_e,t) dT_e,
  \label{line}
\end{equation}
where $T_e$ and $N_e$ are the electron temperature and number density and $G(\lambda_i,T_e,N_e)$ is computed with the CHIANTI V.8 atomic database. To find $G(\lambda_i,T_e,N_e)$  we summed up contributions from each CHIANTI emission line within the $\pm1$\AA{} centered on the wavelength of each individual EVE line, assuming coronal abundances \citep{Feldman1992} and \citet{Mazzotta1998} ionization equilibria.  
Since EVE measures the irradiance in W~m$^{-2}$nm$^{-1}$ and $G(T)$ has units of erg~cm$^{-3}$~s$^{-1}$, to compare between the two we need to take into account the solid angle of the taken solar emission. 
Therefore in Eq.~\ref{line} we divide the EVE irradiances, $\Phi(\mathrm{\lambda_i,t})$, by the approximate solid angle occupied by the area of the AR: $A_\mathrm{AR}/R^2=0.1 A_\mathrm{disk}/R^2=0.1\times 6.807\times 10^{-5}$~sr, where $A_\mathrm{disk}$ is the solar disk area in cm$^2$, R is the Earth-Sun distance and $0.1$ approximates the upper limit of the fraction of the solar disk occupied by the AR \citep[see, e.g.,][]{Schonfeld2017}.  
In Eq.~\ref{line}, $DEM(T_e,t)$ (in units of cm$^{-3}$K$^{-1}$), describes the temperature distribution of plasma emitted along a distance along the LOS, $s$, at temperature $T_e$ at time t,
\begin{equation}
DEM(T_e,t)=N_e^2 ds/dT_e.
\label{dem-eq}
\end{equation}

In this paper we use EVE irradiances, $\Delta \Phi(\mathrm{\lambda_i,t})$, to solve Eq.~\ref{line} and derive $DEM(T_e,t)$.
Previously, many different DEM inversion methods have been developed and used to solve Eq.~\ref{line} for DEM in flares and ARs using AIA/SDO (e.g. \citealt{Cheung2015,Plowman2013,Hannah2012,Warren2012}), EIS/Hinode (e.g. \citealt{Tripathi2011,Warren2012,Testa2011,Petralia2014}), Coronas-F \citep{Reva2018}, EVE/SDO \citep{Schonfeld2017} and EVE/SDO and RHESSI \citep{Caspi2014,McTiernan2019} datasets. 
In this paper we use the \citet{Hannah2012} inversion method with the EVE irradiances, $\Delta \Phi(\mathrm{\lambda_i,t})$, in either the full $23$-line set or on a subset of $5$ lines as identified in Table~\ref{tab:line_power}, and the $G(\lambda_i,T_e,N_e)$ functions from CHIANTI, to find the $DEM(T_e,t)$ (Eq.~\ref{line}). 
It is common to enforce positivity in the DEM solutions to prevent apparently nonphysical negative emission measures, but this may be mathematically incorrect for a difference spectrum in which negative values may occur naturally.
For DEM inversions using the reduced set of $5$ lines we find that the solutions obtained are uniformly positive even without this constraint, which is not the case when we use the full $23$-line set in Table~\ref{tab:line_power}.

Finally, at each time $t_i$, we integrate over all coronal emission lines and DEM($T_e,t$) to find the energy radiated by the AR plasma, the AR luminosity,
\begin{equation}
L_\mathrm{AR}(t)=\int G_\mathrm{tot}(T_e,N_e) DEM (T_e,t) dT_e.
  \label{irrad}
\end{equation}
In this approach, to account for all coronal emission lines in $1-300$~\AA{} range we sum CHIANTI contribution functions, $G_\mathrm{tot}(T_e,N_e)=\sum_{i} G(\lambda_\mathrm{i},T_e,N_e) d\lambda$ using coronal abundances \citep{Feldman1992}, ionization equilibria \citep{Mazzotta1998}, and a constant pressure ($10^{15}$ cm$^{-3}\ K$) within a temperature range of $Log_\mathrm{10}T=[5.4,7.5]$.
%
%
Finally, integrating AR luminosity over time, we find the total AR radiative losses during $10$ days of AR evolution from $t_\mathrm{AR}$ to $t_\mathrm{QS}$.
\begin{equation}
E_\mathrm{rad}(t)=\int_{t_\mathrm{AR}}^{t_{QS}} L_\mathrm{AR}(t)dt.
  \label{loss}
\end{equation}

\section{Results}\label{results}
%

 \begin{figure*}[htb!]
  \centering 
\resizebox{0.95\hsize}{!}
{\includegraphics[angle=0]{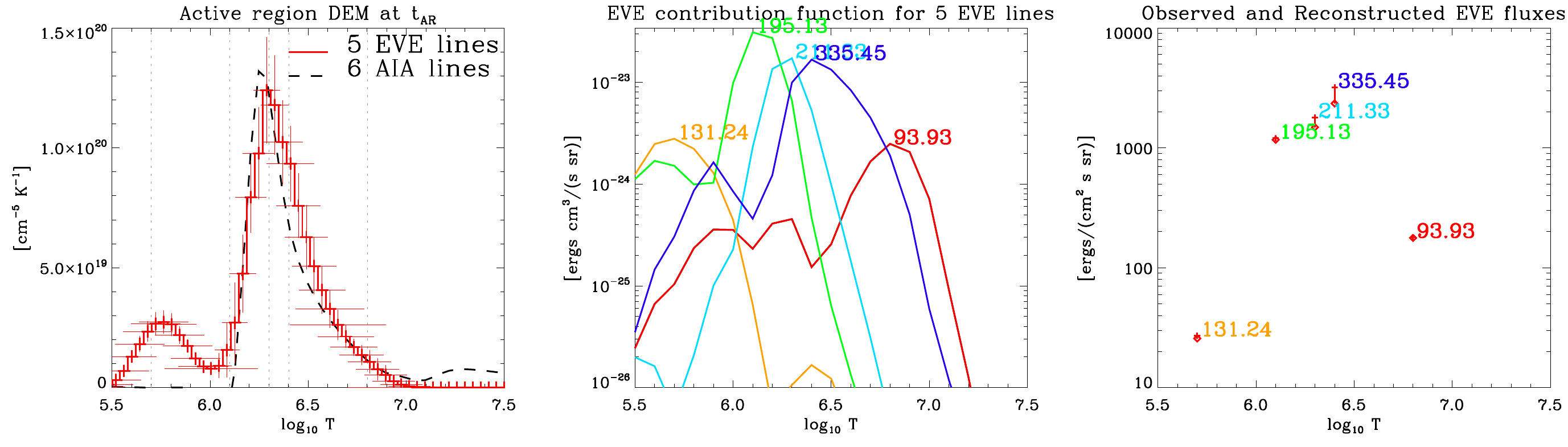}}
 \caption{{\it Left:} Calculated AR DEM using $5$ EVE lines (red) and $6$ AIA bands (dashed black) at $t_\mathrm{AR}$=10-Jul 10:59~UT. {\it Middle:} CHIANTI contribution functions, $G(\lambda_i,T_e,N_e)$, for $5$ selected EVE lines. To find them we summed up contribution functions from each CHIANTI emission line within $\pm1$\AA{} from the line center (see \S\ref{methods}). {\it Right:} EVE irradiances,  $\Delta \Phi(\mathrm{\lambda_i,t_\mathrm{AR}})$, observed and reconstructed from the DEM. See \S\ref{eve_vs_aia}.} 
  \label{fiepsilon_6linesdem}
\end{figure*}

 \begin{figure*}[htb!]
  \centering 
\resizebox{0.95\hsize}{!}{\includegraphics[angle=0]{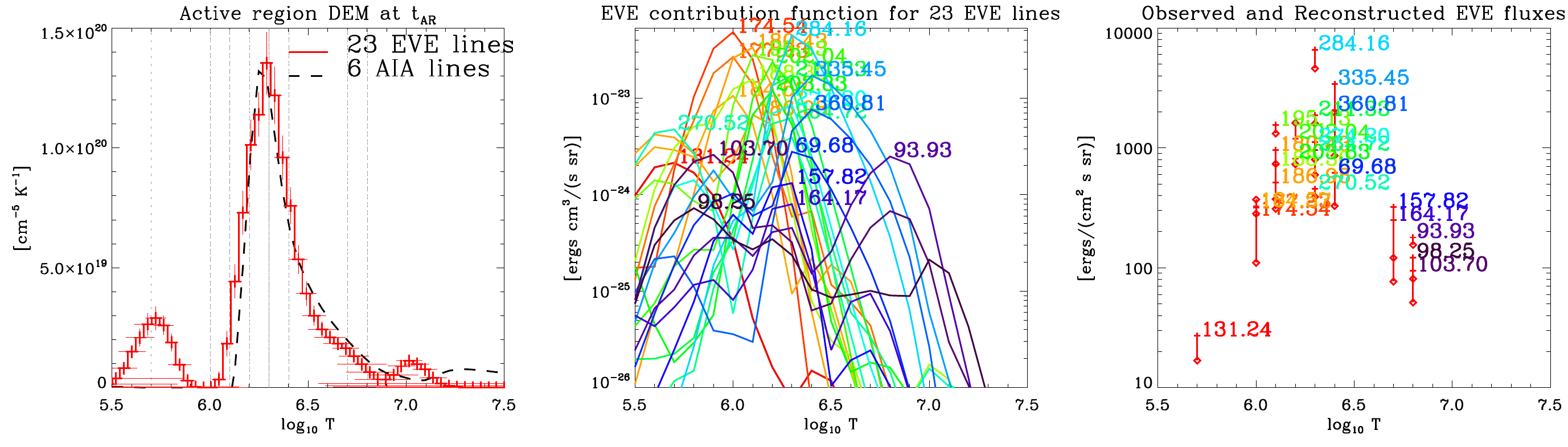}}
 \caption{{\it Left:} Calculated AR DEM using the $23$ EVE lines (red) and $6$ AIA bands (dashed black) at $t_\mathrm{AR}$=10-Jul 10:59~UT. {\it Middle:} CHIANTI contribution functions for the $23$  lines. {\it Right:} EVE irradiances,  $\Delta \Phi(\mathrm{\lambda_i,t_\mathrm{AR}})$, observed and reconstructed from the DEM. See \S\ref{eve_vs_aia}.} 
 \label{dem23-eps}
\end{figure*}

\subsection{Active-Region DEM: EVE vs. AIA observations}\label{eve_vs_aia}
To test our approach of using EVE measurements to find the DEM, we first calculate a DEM from the EVE difference spectrum and compare it with one derived from the full-disk AIA images (Figure~\ref{fiepsilon_6linesdem}). 
To make a direct comparison we first derive the DEM from EVE data using the reduced set of $5$ spectral lines: $131.2$, $195.1$~\AA{}, $211.3$~\AA{}, $335.5$~\AA{}, and $93.9$~\AA{}. 
We exclude the $171.1$~\AA{} line from our analysis since its difference irradiance, $\Delta 
\Phi(\mathrm{\lambda_i,t_\mathrm{AR}})$, dropped below zero as AR 11520 crossed the disk (see Figure~\ref{linefits_evol}, Panel 2 for $171.1$~\AA{}). For AIA intensities we use full-disk averaged intensity differences, $\bar{I}_\mathrm{AIA,\lambda}(t_{AR})-\bar{I}_\mathrm{AIA,\lambda}(t_{QS})$. 
For AIA emissivities we use the AIA response function \citep{Boerner2012}. 
To find the DEM we choose a range of temperatures, $Log_\mathrm{10}T=[5.4,7.5]$, binned into 53 segments.  

Figure~\ref{fiepsilon_6linesdem} shows an example of two DEMs, $DEM(T_e,t_\mathrm{AR})$, derived from difference fluxes from EVE ($5$ lines) and AIA ($6$ bands), $\Delta \Phi(\mathrm{\lambda_i,t_\mathrm{AR}})$ (see Eq.~\ref{eq_flux}), respectively. 
The middle panel shows EVE contribution functions corresponding to each line. Notice that even though many lines can be considered isothermal, they have significant emission over a range of temperatures due to contributions from other lines within $\pm1$\AA{} centered on each individual EVE line (see $G(\lambda,T_e,N_e)$ description in \S\ref{methods}).
The right panel shows the comparison between observed and DEM-reconstructed EVE fluxes, as a validation test for the DEM inversion. 
From this comparison, we conclude that the EVE and AIA datasets result in similar DEM estimates, validating our working hypothesis of using EVE Sun-as-a-star measurements to derive thermal properties of an evolving AR in the absence of other solar activity on the disk.
The coarse wavelength set for EVE resulted in a minor discrepancy at the lowest temperatures.

In Figure~\ref{dem23-eps} we also calculate the ARÕs DEM when it is at the disk center using all $23$ EVE lines instead of the $5$-line subset as above.  
We compare DEMs from $23$ EVE lines with DEM derived using $6$ bands from AIA and also find that the two agree well,
but note that the errors for the $23$-line inversion are larger (compare e.g. the $131$~\AA{} and  $94$~\AA{} lines validation on the right panels of Figures ~\ref{fiepsilon_6linesdem} and \ref{dem23-eps}). These larger errors might lead to non-realistic DEM inversions (see e.g. a small DEM increase around $Log_\mathrm{10}T=7$ in the $23$-line inversion). Moreover  the selection of suitable lines for DEM analysis is critical since the detailed atomic characteristics associated with the chosen emission lines must be fully characterized to properly calculate the DEM. 
Since CHIANTI lacks a large number of weak unresolved lines that appear in the EVE spectrum, we further restrict our analysis of DEM evolution to only $5$ emission lines, shown in Figure~\ref{fiepsilon_6linesdem}, that are best characterized.

\begin{figure*}[htb!]
  \centering 
\resizebox{0.35\hsize}{!}{\includegraphics[angle=0]{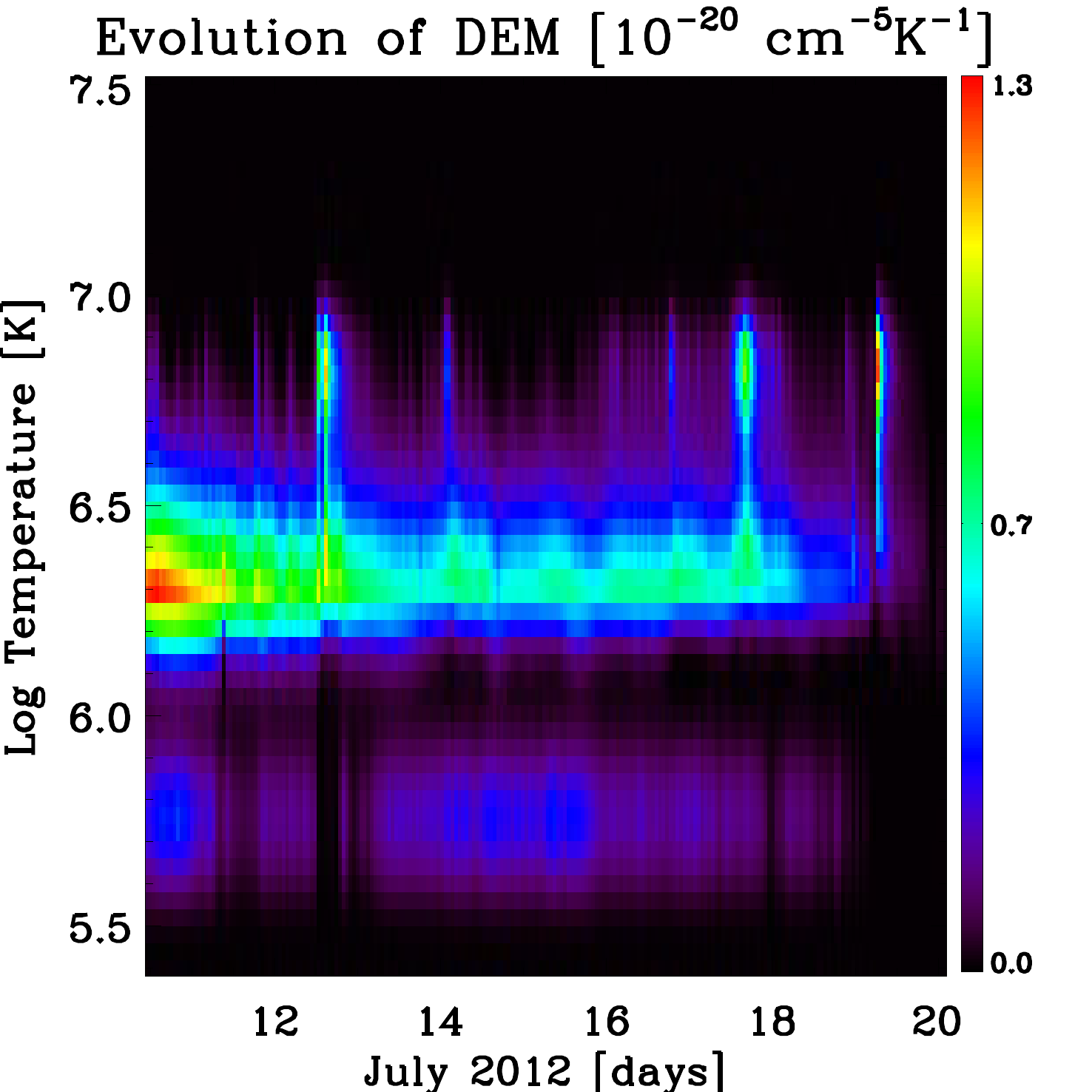}}
\resizebox{0.64\hsize}{!}{\includegraphics[angle=0]{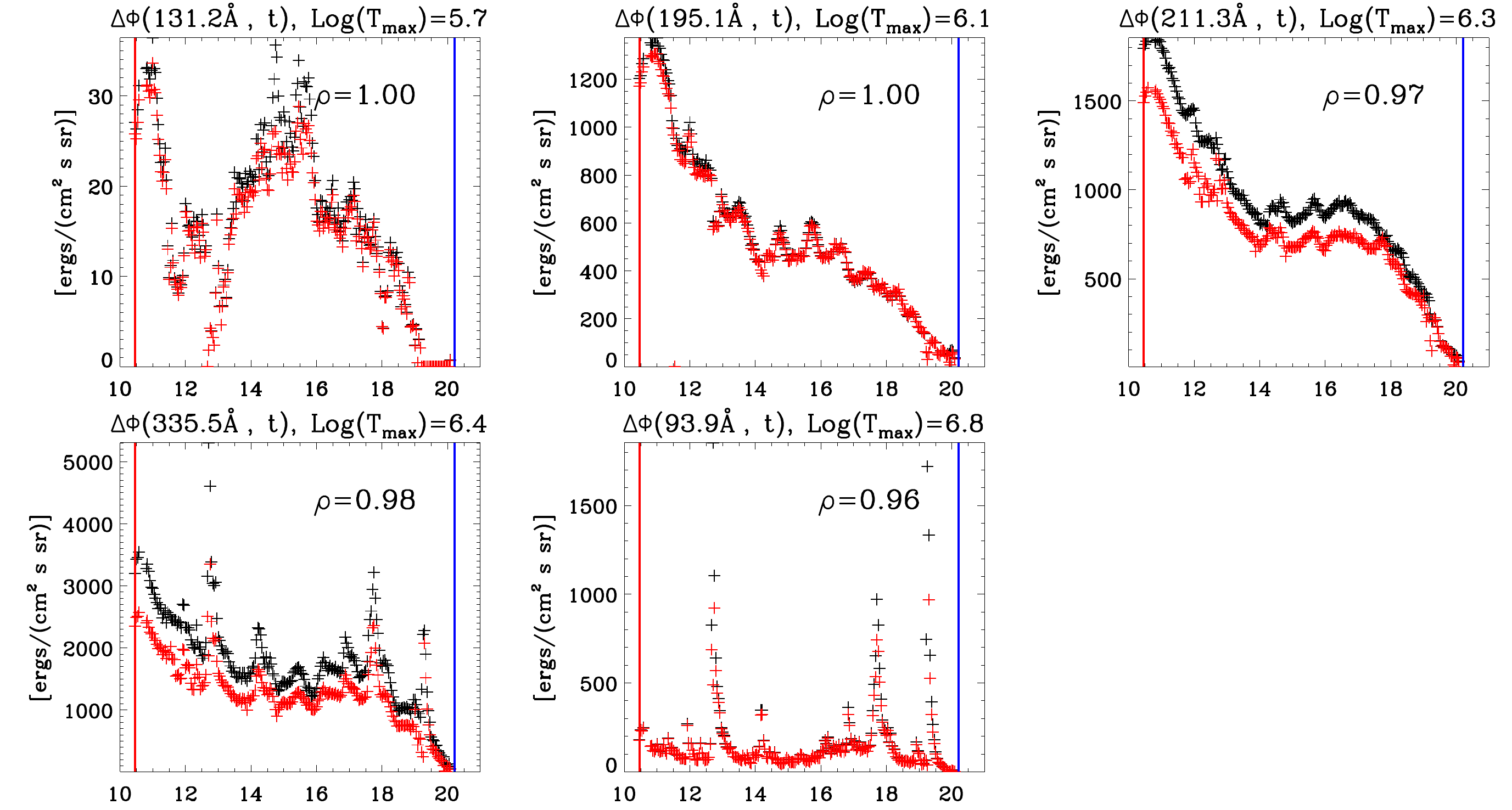}}
 \caption{ {\it Left:} The evolution of the AR 11520  $DEM(t,T_e)$, represented as a a spectrogram.
 Note the occasional extensions to higher temperatures associated with flares, such as on July 12 2012 when an X1.4-flare occurred.
 {\it Right:} Quality of the DEM reconstruction: observed EVE input fluxes (black) and fluxes reconstructed from the DEM (red) vs. time. $\rho$ in the top right corner shows a Spearman correlation coefficient between EVE input and DEM-reconstructed output fluxes. See \S\ref{irrad_evol}.}
  \label{fiepsilon_dem}
\end{figure*}

 \begin{figure}[htb!]
  \centering 
\resizebox{1.0\hsize}{!}{\includegraphics[angle=0]{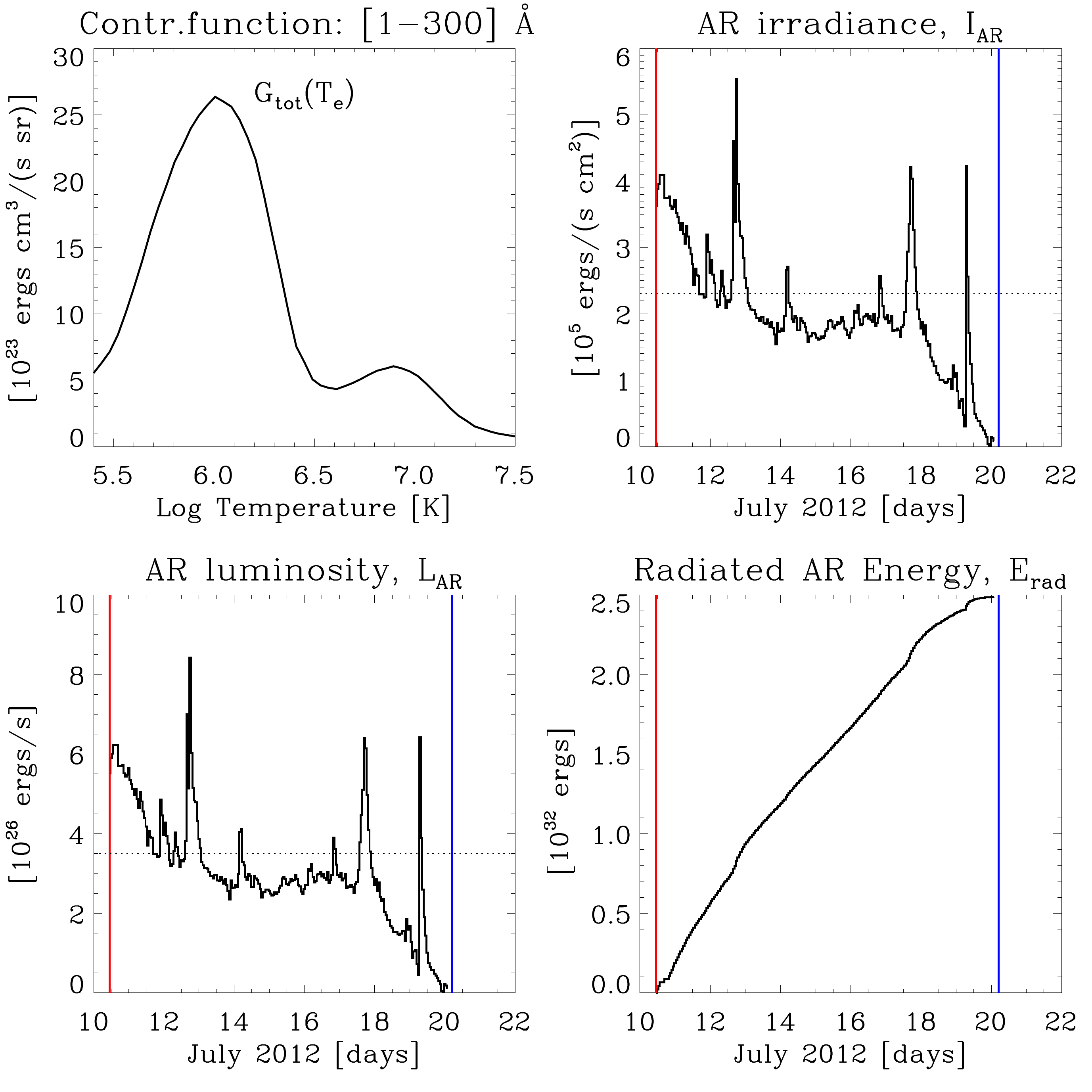}} 
 \caption{{\it Top row:} Total CHIANTI contribution function, $G_\mathrm{tot}(T_e,N_e)=\sum_{i} G(\lambda_\mathrm{i},T_e,N_e) d\lambda$, in the $1-300$~\AA{} range and derived AR 11520 irradiance,  $I_\mathrm{AR}$; {\it Bottom row:} AR luminosity, $L_\mathrm{AR}$, and cumulative AR radiative losses, $E_\mathrm{rad}$. Horizontal dotted lines show mean values of  $\bar{I}_\mathrm{AR}=6\times10^{6}$~ergs/s and $\bar{L}_\mathrm{AR}=9\times10^{27}$~ergs/s, when AR 11520 was present on the solar disk (from July 10 to July 18). See \S\ref{irrad_evol}.}
\label{lumin}
\end{figure}

\subsection{Active Region Irradiance Evolution}\label{irrad_evol}
 \begin{figure*}[tbh!]
  \centering 
\resizebox{0.45\hsize}{!}{\includegraphics[angle=0]{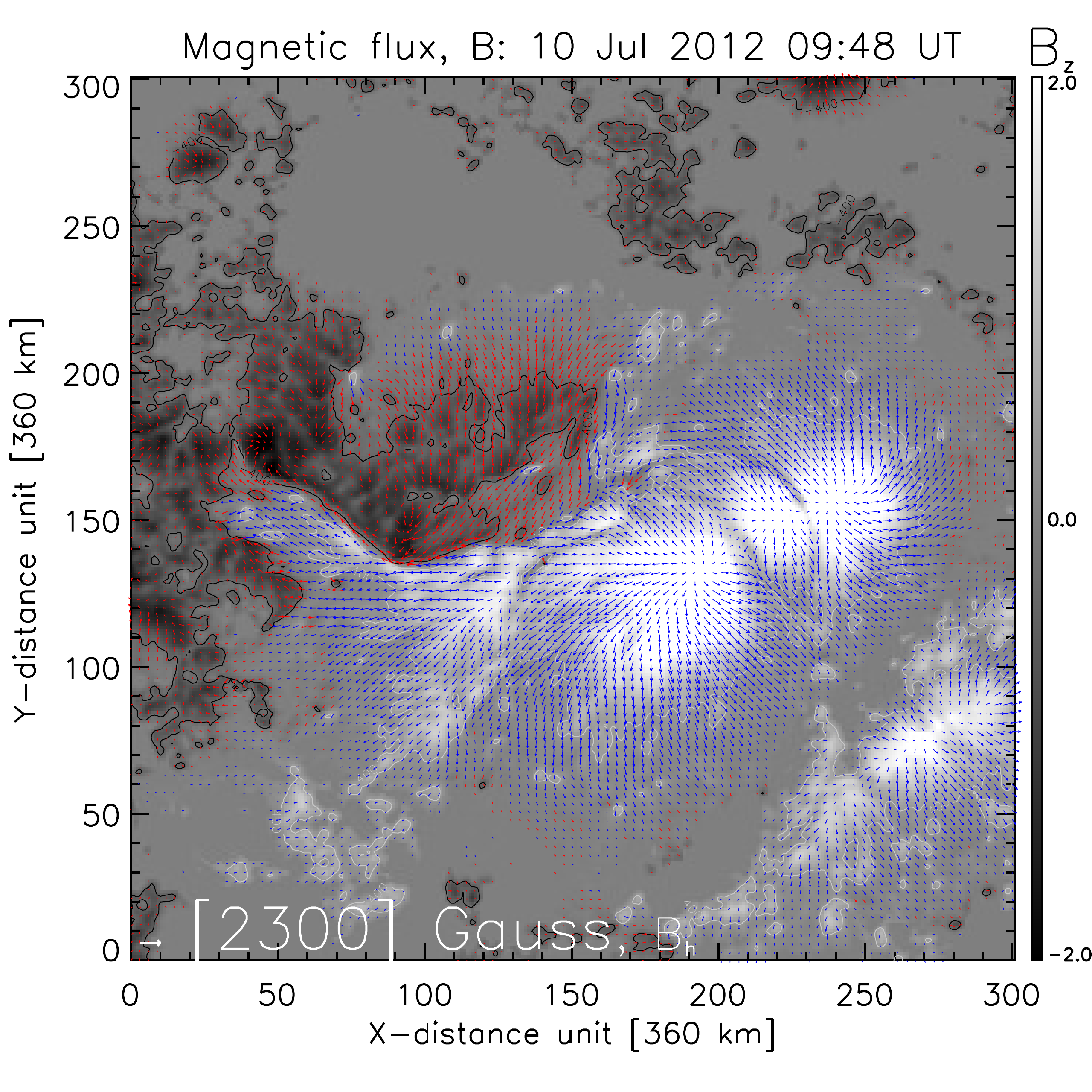}} 
\resizebox{0.45\hsize}{!}{\includegraphics[angle=0]{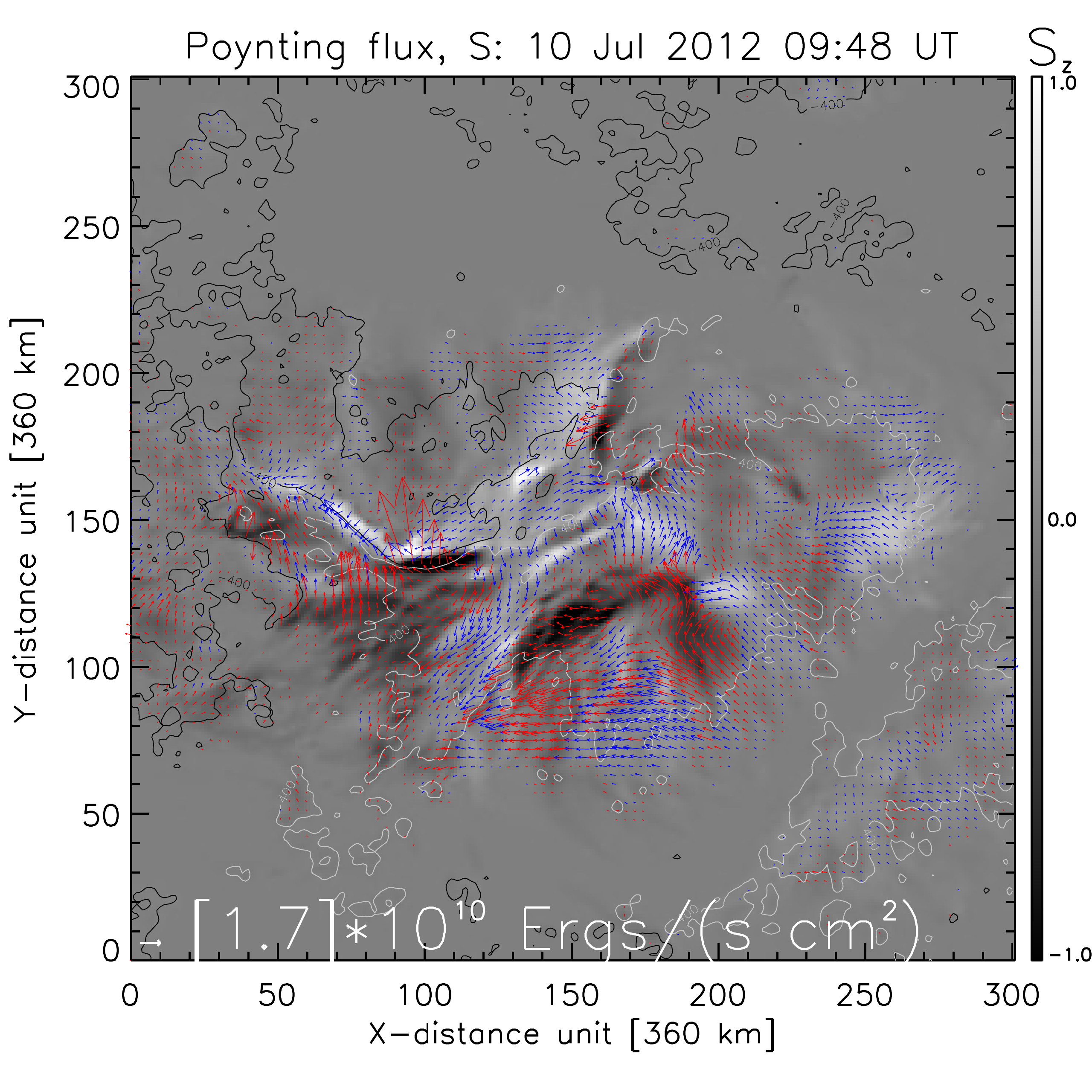}} 
 \caption{ AR 11520 vector magnetic field {(\it Left)} and the PDFI energy flux vector through the photosphere {(\it Right)} on July 10 2012, 09:48~UT. The $B_z$ and $S_z$ black-and-white color scales vary from -2000 to 2000 G and $-10^{10}$ to  $10^{10}$ ergs/(cm$^2$~s), respectively. See \S\ref{irrad_evol}.}
  \label{figsz}
\end{figure*}

Figure~\ref{fiepsilon_dem} (left panel) shows the evolution of the derived AR $DEM (T_e,t)$ from July 10 to July 20 2012. 
The panels on the right show the quality of the DEM reconstruction: a comparison between EVE input difference fluxes and the fluxes reconstructed from the DEM. 
The validation time series match each other closely, including the hot enhancements during flares, with Spearman correlation coefficient ranging from 0.96 to 1.0 for the $5$ selected lines. 
We find the worst differences (up to $20\%$ for $211$~\AA{} and $30\%$ for $335$~\AA{} lines), corresponding to $Log_\mathrm{10}(T_\mathrm{max})~=~[6.3,\ 6.4]$.
These discrepancies serve as uncertainty estimates, complementing contributions from the calibration of the EVE irradiance spectra ($5-7\%$, for the strongest lines, \citealt{Hock2012}), error estimation of line fluxes from our Gaussian fits to the EVE spectra (several percent depending on the line), any density sensitivity in the DEM estimates, plus the acknowledged uncertainties in the CHIANTI atomic data. 
We estimate an overall uncertainty in the DEMs to be less than $40\%$, exclusive of the possibly comparable uncertainties in the atomic data \citep{Dere1997}.

In Figure~\ref{lumin} we show how the EUV irradiance and luminosity evolve during the $10$ days of AR evolution, $I_\mathrm{AR}(t)$ and $L_\mathrm{AR}(t)$. 
To estimate $L_\mathrm{AR}$(t) (top right panel), we use Eq.~\ref{irrad}, using the EVE-derived $DEM(T_e,t)$ (see Figure~\ref{fiepsilon_dem}) and the total contribution function, $G_\mathrm{tot}(T_e,N_e)=\sum_{i} G(\lambda_\mathrm{i},T_e,N_e) d\lambda$, using CHIANTI data in the $1-300$~\AA{} range 
(top left panel). 
We find the mean AR 11520 irradiance to be $\bar{I}_\mathrm{AR}=2.3\times10^5$~ergs/(cm$^2$~s), and the luminosity to be $\bar{L}_\mathrm{AR}=3.5\times10^{26}$~ergs/s~=~$10^{26.5}$~ergs/s (see horizontal dotted lines on the top right and bottom left panels, respectively).  
Integrating $L_\mathrm{AR}$ over $10$ days, we find the total radiative energy losses of AR 11520 during the 10 days of AR evolution to be $E_\mathrm{rad}=2.5\times10^{32}$~ergs in the 1--300~\AA{} range.

For comparison, the range of solar cycle variation of the luminosity of the Sun in this spectral range has been estimated at $3\times10^{25}$~ergs/s to $10^{27}$~ergs/s \citep{Hunsch1998,Hunsch1999,Haisch1996}.  
\citet{Acton1996} used \textit{Yohkoh}/SXT data converted to the  ``RASS'' $[0.1-2.4]$~keV ($[5.2-123]$~\AA{}) ROSAT/PSPC passband to find similar estimates; these were later adjusted by \citet{Haisch1996} to increase the minimum value to $10^{26}$~ergs/s.
\citet{Ayres1997} used suborbital rocket measurements in a narrower passband, $[6-62]$~\AA~($0.2-2$~keV), to obtain higher variations from $5\times10^{26}$ to $2\times10^{27}$~ergs/s. 
Finally, \citet{Judge2003} also used SNOE-SXP measurements in the RASS passband to estimate variations from $6.3\times10^{26}$ to $7.9\times10^{27}$~ergs/s.
To relate our estimates of AR luminosity with stellar and solar cycle variations, we also evaluate radiative loss rates over the RASS passband.  We find the mean AR 11520 luminosity, $\bar{L}_\mathrm{AR,(3-124)}=1.4\times10^{26}$~ergs/s, which is some 5 times smaller than the minimum range of the solar cycle variation, $6.3\times10^{26}$~ergs/s, reported by \citet{Judge2003}, and therefore roughly in concordance given the common occurrence of multiple active regions during solar maximum.

In addition, we compare our region's luminosity to the magnetic energy flux through the photosphere, the Poynting flux $S_z$, one hour before $t_\mathrm{AR}$. Figure~\ref{figsz} shows the spatial distribution of $S_z$ in AR 11520.  
To calculate S$_\mathrm{z}=-{\bf E_h} \times {\bf B_h}$ we use the horizontal component of the electric field vector, ${\bf E_h}$, from the PDFI  inversion method \citep{Kazachenko2014,Kazachenko2015}, and the horizontal component of the vector magnetic field ${\bf B_h}$, on July 10 09:48~UT. 
We find that the majority of the AR magnetic energy at the photosphere is injected at positive and negative vertical Poynting fluxes, $S_z\approx[10^7,10^8]$~ergs/(cm$^2$~s), with a total AR-integrated energy injection rate, $E_\mathrm{mag}(t_\mathrm{AR})=\int_\mathrm{AR} S_z dS=3\times10^{28}$~ ergs/s. 
Note, that here we only estimate the energy injection rate $E_\mathrm{mag}$ at one time, when the AR was close to the disk center. 
However, as we have seen from a study of magnetic energy fluxes in AR 11158, $E_\mathrm{mag}(t)$ varies significantly as a function of time.
For AR 11158 we found a mean $\bar{E}_\mathrm{mag}=5\times 10^{27}$~ergs/s during 5 days of this region's evolution, varying from $-6$ to 18$\times10^{27}$~ergs/s (see Fig.~9 in \citealt{Kazachenko2015}). 
If we assume, that our instantaneous estimates of $E_\mathrm{mag}(t_\mathrm{AR})$ and $S_z(t_\mathrm{AR})$ approximate mean magnetic energy injection rate at the photosphere, as AR 11520 evolved, then we conclude that both typical Poynting fluxes, $S_z(t_\mathrm{AR})\approx[10^7,10^8]$~ergs/(cm$^2$~s), and the mean magnetic energy injection rate, $E_\mathrm{mag}(t_\mathrm{AR})=3\times10^{28}$~ergs/s, are around two orders of magnitude larger than the mean AR irradiance,  $\bar{I}_\mathrm{AR}=2.3\times10^5$~ergs/(cm$^2$~s), and the AR luminosity, $\bar{L}_\mathrm{AR}=3.5\times10^{26}$~ergs/s, derived above. 

Finally, we compare our mean AR luminosity to AR unsigned magnetic flux, $\bar{L}_\mathrm{AR}$  and $\Phi_\mathrm{LOS,AR}$. Previously, \citet{Fisher1998} used Mees Solar Observatory (MSO) vector magnetic field and \textit{Yohkoh} SXT X-ray observations in $333$ ARs between 1991 and 1995 to find a relationship between magnetic fields and coronal heating. 
They deduced a scaling law between AR luminosities (in $1-300$~\AA{}) and unsigned magnetic fluxes, $L_{AR}\approx1.2\times10^{26}\ \mathrm{erg/s}\ (\Phi_\mathrm{AR}/10^{22}$Mx $)^{1.19}$.  
Note that \citet{Fisher1998} used a simplified single-temperature approach ($Log_\mathrm{10}T=6.5$ or $T=3\times10^6$~K) resulting in errors in $L_\mathrm{AR}$ up to a factor of $2$. 
Here, in contrast, we use a more accurate multi-temperature DEM approach to estimate the AR luminosity. 
We compare our estimates of mean AR luminosity and magnetic flux as AR crossed the disk, $\bar{L}_\mathrm{AR}=3.5\times10^{26}$~erg/s and $\Phi_\mathrm{LOS,AR}=8\times10^{22}$~Mx. 
We find that our $\bar{L}_\mathrm{AR}$ is $4$ times smaller than the one expected for a given $\Phi_\mathrm{LOS,AR}$ from the scaling law above, but is in agreement with the overall scatter of $(L_{AR}-\Phi_\mathrm{AR})$ shown in Figure 9 of \citet{Fisher1998}. 
Note that here we only compared mean values of $(L_{AR}-\Phi_\mathrm{AR})$, instead of their evolution, since neither of these variables exhibited much variability apart from variations due to solar rotation. 
We would also like to note that our AR magnetic flux is above the range of AR magnetic fluxes in \citet{Fisher1998}. 
Comparing unsigned magnetic fluxes in \citet{Fisher1998} from MSO and \citet{Kazachenko2017} from SDO, we find that the SDO values are around $5$-times larger than the MSO values, a discrepancy that might be interesting to explore in future.

\section{Conclusions}\label{discussion}

We have evaluated the radiative losses during a quiescent period of the evolution of AR 11520. 
For this we used hourly Sun-as-a-star spectra from EVE/SDO in the $6.5$~nm to $37$~nm range during $10$ days, as this 
fairly large AR 11520 ($\Phi_\mathrm{LOS,AR}=8\times10^{22}$~Mx) was crossing the solar disk. 
Since AR 11520 was the only major region on the solar disk, we could use EVE difference fluxes, 
$\Delta \Phi(\mathrm{\lambda_i,t}) =\Phi(\lambda_i,t)-\Phi(\lambda_i,t_\mathrm{QS})$, to derive its DEM.  
Specifically, we used a set of $5$ iron emission lines that sample the AR coronal temperature range from 0.5~MK to $6.3$~MK ($Log_\mathrm{10}T=[5.7,6.8]$) and have well-studied atomic properties and minimal blending.
We then applied this approach to a sequence of EVE spectra to derive a sequence of DEMs during $10$ days of AR evolution (see Figure~\ref{fiepsilon_dem}). 
From the derived evolution of DEM we determined the AR radiative losses as a function of time (see Figure~\ref{lumin}). 
Our findings are as follows.
\begin{enumerate}
 \item We find that the shape and the peak temperature of the DEM, as extracted from EVE data,
 are overall consistent with previous studies using spatially-resolved studies of quiescent regions (e.g. \citealt{Warren2012,Petralia2014,Tripathi2011}). 
 Our DEM peaks at  $Log_\mathrm{10}T=6.3$, corresponding to a relatively cool AR plasma. 
 The peak temperature stays nearly constant as the AR evolves, illustrating the slowly varying nature of irradiance of the region studied. 
 The DEM does exhibit high-temperature enhancements during flares. 
Our 1-hour cadence of EVE data is too sparse for describing flare contributions, and we have ignored them in this analysis.


 \item We find the mean AR irradiance during 10 days of AR evolution in the $1-300$~\AA{} spectral range, ${\bar I_\mathrm{AR}}=2.3\times10^{5}$~ergs/(cm$^2$s). 
 This estimate is around $20$ times smaller than the classical \citet{Withbroe1977} estimate of the bolometric radiative flux in such a region, $I_\mathrm{AR,W\&N}=5\times10^{6}$~erg/(cm$^2$s). Note, however, that our estimate does not include the optically thick component (essentially, the chromosphere), which \citet{Withbroe1977} estimate at about twice the magnitude of the optically thin coronal component.
 The remainder of the discrepancy presumably relates to the scaling of AR area.
 
 \item Integrating over the whole AR, we find a mean AR radiative energy loss rate during 10 days of AR evolution, ${\bar L_\mathrm{AR}}=3.5\times10^{26}$~erg/s.
We compare it with the magnetic energy flux through the photosphere, when the AR is close to the disk center,  $S_\mathrm{z,mag}=[10^7,10^8]$~erg/(cm$^2$s), or integrating over the whole active region, a total energy flux of $E_\mathrm{mag}=3\times10^{28}$~erg/s. 
We conclude that the coronal radiative energy losses within $1-300$~\AA{} comprise about 1\% of the available magnetic energy flux, confirming the relative importance
of the chromospheric heating problem.

\item To relate our estimates of AR luminosity with stellar and solar cycle variations, we evaluate radiative loss rates over the $3-124$~\AA{} ROSAT-PSPC passband. We find a mean AR 11520 luminosity, $\bar{L}_\mathrm{AR,(3-124)}=1.4\times10^{26}$~erg/s, which is $5$ times smaller than the minimum range of the solar cycle variation, $6.3\times10^{26}$~erg/s, reported by \citet{Judge2003}. 

 \item We find AR 11520 total cumulative radiated energy,  $E_\mathrm{rad}=2.5\times10^{32}$~ergs, during 10 days of evolution. This estimate is an order of magnitude larger or similar to bolometric radiated energies associated with M- and X-class flares  (\citealt{Emslie2012}, $E_\mathrm{bol}\approx10^{30}-10^{31}$ ergs), highlighting importance of AR radiative losses in AR and flare energetics. 
 
\item Finally, we compare our mean AR luminosity to AR unsigned magnetic flux. We find these to be consistent with a previously derived statistical relationship,  $L_{AR}\sim\Phi_\mathrm{AR}^{1.19}$, using a more simplified single-temperature approach \citep{Fisher1998}. We therefore confirm that this relationship could be further used in both solar and stellar astronomy, when only one kind of observation is available. 

\item This study is the first detailed analysis of AR thermal properties using EVE/SDO Sun-as-a-star observations. 
Our novel approach opens doors to similar studies of active regions on other stars where spatial resolution is an issue (e.g. \citealt{France2019}).

\end{enumerate}

To conclude, our analysis demonstrates that AR coronal radiative losses are $\sim 100$ times smaller than the typical magnetic energy fluxes at the photosphere, but nevertheless are significant compared to flare radiative losses. 
Although this study did not include chromospheric losses, \cite{Withbroe1977} estimate these at about twice the coronal contribution, and the majority of the magnetic energy supply may therefore wind up in faculae.
Future synergetic studies of energy cycle in active regions, including magnetic energy fluxes through the photosphere and coronal energy losses during both quiescent and active periods of AR evolution (e.g. \citealt{Iglesias2019}) will enable our understanding of active-region energetics that may be overlooked in isolated studies of flaring or quiescent periods of ARs. 

\acknowledgments
We thank Richard C. Canfield for illuminating discussions that have triggered this project. We thank the EVE and AIA teams for providing us with the SDO/EVE and SDO/AIA datasets. We thank the US taxpayers for providing the funding that made this research possible. We acknowledge support from NASA LWS NNH17ZDA001N (M.D.K), NASA 80NSSC18K1283-HSR (M.D.K) and NASA ECIP NNH18ZDA001N (M.D.K.). 
H.S.H. thanks the University of Glasgow for hospitality.
\bibliography{full_lib_thes_cp}

\end{document}